# Optical wavefront phase-tilt measurement using Si-photonic waveguide grating couplers

S. Janz,[1,*] D.-X. Xu,[1] Y. Grinberg,[2] S. Wang,[1] M. Vachon,[1] P. Cheben,[1] J.H. Schmid,[1] and D. Melati[3]

[1]*Advanced Electronics and Photonics Research Centre, National Research Council Canada, Ottawa, Canada*
[2]*Digital Technologies Research Centre, National Research Council Canada, Ottawa, Canada*
[3]*Centre for Nanoscience and Nanotechnology, Université Paris Saclay, CNRS, 91120 Palaiseau, France*

*Corresponding author: siegfried.janz@nrc-cnrc.gc.ca*

**Abstract:** Silicon photonic wavefront phase-tilt sensors for wavefront monitoring using surface coupling grating arrays are demonstrated. The first design employs the intrinsic angle dependence of the grating coupling efficiency to determine local wavefront tilt, with a measured sensitivity of 7 dB/°. A second design connects four gratings in an interferometric waveguide circuit to determine incident wavefront phase variation across the sensor area. In this device, one fringe spacing corresponds to approximately 2° wavefront tilt change. These sensor elements can sample a wavefront incident on the chip surface without the use of bulk optic elements, fiber arrays, or imaging arrays. Both sensor elements are less than 60 μm across, and can be combined into larger arrays to monitor wavefront tilt and distortion across an image or pupil plane in adaptive optics systems for free space optical communications, astronomy and beam pointing applications.

In free-space optical communications, atmospheric turbulence limits the signal power that can be focused on a detector, adds phase and amplitude distortion to the signal and causes signal drop out. Turbulence also severely reduces the image resolution of large astronomical telescopes. The astronomy community has developed adaptive optics correction systems to address this problem [1-3], and ground based observatories now achieve image resolution comparable to space telescopes, which do not suffer from atmospheric turbulence. An adaptive optics system continuously measures the distortion of an incoming wavefront and then corrects the wavefront incident at the image plane or receiver by using, for example, a deformable mirror. The initial wavefront measurement step may be done using a Shack-Hartmann wavefront monitor comprising an array of hundreds of lenses that focus the pupil plane onto an imaging array [1,2]. Other established wavefront monitor systems also involve complex optical assemblies often built on an optical bench. This is acceptable for large one-of-kind telescopes that use customized adaptive optics systems. In free-space communications and also other medical or industrial applications involving light transmission through optically distorting and time varying media, very small and fast wavefront sensors are needed. Integrated photonics has the capability to do complex on-chip optical processing on monolithic millimeter size chips. To date there has been some work on hybrid assemblies of fiber optic arrays and beam combiner chips for wavefront sensing [3]. However, there has been little work on a purely integrated optic approach to wavefront sensing. In this work we demonstrate the use of silicon photonic waveguide circuits with surface grating couplers to directly sample the light incident on the chip surface, as shown schematically in Fig. 1(a), and determine the local wavefront tilt. Two approaches are explored and demonstrated in this work. The coupling ratio sensor employs the intrinsic angle dependence of the grating coupling efficiency to measure the local incident wavefront angle. The second approach uses on-chip interferometry to measure the relative incident wavefront phase at different points on the chip surface. Both of these tilt sensors can be combined into arrays of sensors to form a wavefront monitor that tracks more complex wavefront distortions over a wide area.

The coupling ratio sensor in Fig. 1(b) is formed by a pair of surface grating couplers with a center-to-center separation of 35 μm. Each grating is designed to couple light at near normal incidence on the chip into a waveguide. In this device, the peak coupling angle is chosen to be a few degrees away from exact normal incidence. When the gratings are oriented in opposite directions as in Fig. 1(a), the ratio of powers coupled into the two output waveguides R1 and R2 wave will vary with local wavefront tilt, because of the angular dependence of the grating coupling efficiency. For closely spaced gratings the light intensities striking each grating are assumed to be equal, as would be the case when sampling a wavefront at the imaged pupil plane of a telescope.

The interferometric wavefront sensor layout is shown in Fig. 1(c). This sensor element consists of four grating couplers arranged with center-to-center separations of 35 μm and 21 μm along x- and y-directions, respectively. In this design, the peak grating coupling angle into output waveguides is preferably at normal incidence. The incident light coupled into the waveguides by each grating is combined with light coupled from the adjacent gratings by Y-junction combiners, which launch the four output signals into the output waveguides F1 to F4. The output signal powers are determined by the relative phase of the light incident on two connected gratings. The four outputs therefore provide information on the relative phases of the incident wavefront at each sensor element across the array. The waveguide signals from both the devices in Fig. 1 were coupled from the chip to single mode fibers by coupling gratings located near the edge of the chip, and directed to a photodetector.

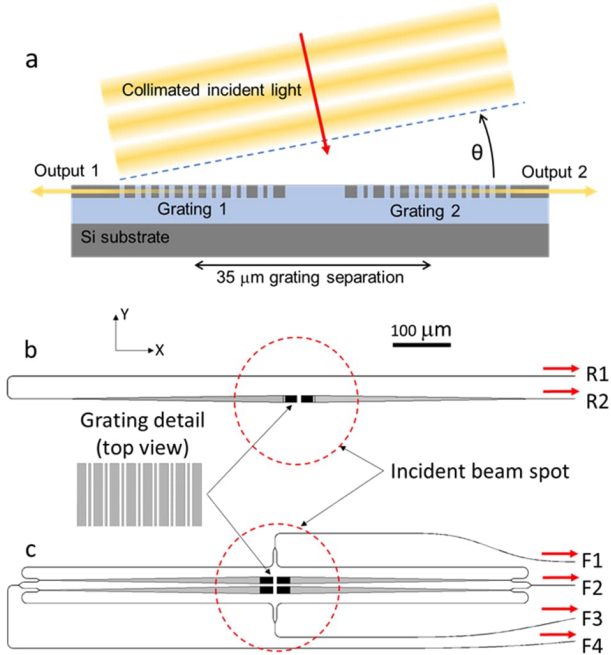

Fig. 1. (a) A schematic side view of the wavefront tilt measurement configuration. The input beam is incident from above at a small wavefront tilt angle θ. Top views of the layout for (b) the coupling ratio sensor and (c) the interferometric sensor. The output signals R1 and R2, and F1 to F4, are coupled off chip to optical fibers and a photodetector (not shown).

The devices were fabricated on a silicon-on-insulator (SOI) wafer with a 220 nm thick Si waveguide layer over a 2 μm buried $SiO_2$ layer, and covered by an $SiO_2$ cladding. The 500 nm wide channel waveguides were formed by etching fully through the Si layer. The incident wavefront is sampled by surface grating couplers with a nominal design peak coupling efficiency angle at normal incidence (θ = 0°) for a wavelength of λ=1550 nm. In the experiments, the peak coupling angle needed to optimize each sensor type was adjusted by changing the incident wavelength, which results in only a small penalty in coupling efficiency. The gratings were designed to couple incident transverse electric (TE) polarized light into the output waveguides. Each grating is 25 μm long and 15 μm wide, and connected to the 500 nm wide Si output waveguides by a 400 μm long adiabatic waveguide taper. The near field output beam of the grating was originally designed to couple light to a single mode fiber with a 10.4 μm mode field diameter. Each period of the grating comprised four parallel 15-μm-long Si and $SiO_2$ strips oriented at 90° to the beam propagation direction, as shown in the inset of Fig. 1. The grating materials (strip width), in sequential order from the waveguide end of the grating, were $SiO_2$ (95 nm), Si (81 nm), $SiO_2$ (114 nm), and Si (358 nm) and were formed by etching the grating trenches completely through the 220-nm-thick Si layer in the same etch step used to form the waveguides. This 648-nm-long unit cell pattern was repeated over the 25-μm-long grating length without apodization. This grating pattern is conceptually similar to that of Ref. [4] but with structure dimensions determined using the machine learning methodology developed by Melati *et al.* [5, 6]. These gratings are designed to suppress the strong back-reflections that would otherwise occur because vertical incidence grating couplers also satisfy the Bragg condition for back-reflection. If not suppressed, these Bragg enhanced back-reflections cause significant distortions in the coupled optical signals [7]. In 2-dimensional FDTD numerical simulation the coupling efficiency to single mode fiber with a 10.4 μm diameter mode was approximately -3 dB, with a 60 nm wide 3dB wavelength bandwidth, and less than -20 dB back-reflection.

The light source was a fiber coupled laser with a wavelength tuning range from 1470 nm to 1580 nm. A TE polarized incident plane wave was formed by a fiber collimator that produced a beam with a $1/e^2$ intensity full-width of approximately 2 mm. The measured incident power variation was less than 3% across a 60 μm diameter area on the chip surface enclosing the wavefront sensor arrays. The collimator was mounted on a goniometer assembly that allowed the incident angle to be varied from θ = -6° to θ = +6° around normal incidence, while keeping the beam centered on the grating array. Sensor performance over a finite wavelength band Δλ was emulated by tuning the laser across the grating bandwidth and averaging the output spectra.

A fabricated test grating was characterized prior to performing wavefront sensor measurements. The coupling efficiency to single mode fiber was approximately -5 dB. The peak free-space beam coupling efficiency at normal incidence θ = 0° was at a wavelength of λ =1515 nm for the test grating, but it was observed that the peak wavelength varied slightly for gratings at different locations on the chip, due to small fabricated dimension variations. The 3 dB coupling bandwidth was approximately 30 nm. The peak coupling angle could be changed to θ =3° by changing the wavelength to approximately λ =1525 nm. Within a few degrees from normal incidence, the measured incident beam to waveguide coupling pattern closely approximates a symmetric Gaussian distribution $A(\theta) \sim \exp(-2\theta^2/\theta_0^2)$, with a FWHM of 4° ($\theta_0$ = 2.4°) along the grating propagation axis and 14° ($\theta_0$ = 8°) perpendicular to the propagation axis. At larger angles the coupling envelope decays more slowly than an ideal Gaussian because the gratings were not apodised. By reciprocity, $A(\theta)$ is the same as the far field intensity pattern that would be radiated by the grating when used as an output coupler. The differences in fabricated grating peak coupling angles and wavelengths from the design targets given above are attributed to changes in the effective index of the grating caused by small differences from the nominal design dimensions of the Si structure, as well as the approximations inherent in the 2D FDTD simulations.

The coupling ratio wavefront sensor was tested by illuminating the two grating sensor elements, as shown in Fig. 1(a), with the collimated TE polarized beam. A 2 nm wide incident wavelength band from λ = 1525 nm to 1527 nm was used, resulting in a peak coupling angle of approximately θ = 3°. For a 1 mW incident beam power the maximum coupled power into each output waveguide was approximately 20 nW. Fig. 3 shows the ratio of powers coupled into the two gratings as the incident wavefront tilt was varied between θ = -5° and θ = 5°. Within a ±2° range near normal incidence the ratio shows a strong variation with tilt angle with a constant logarithmic slope of 7 dB/°, as would be expected for a ratio of two Gaussian coupling envelopes centered at opposite angles on either side of the surface normal axis. At tilt angles beyond ±3° the R1/R2 ratio decreases since the coupling envelope tail does not decay as rapidly as a Gaussian distribution at larger angles. Since the ratio is independent of incident light intensity, the tilt angle measurement is not affected by incident beam power fluctuations. However, the measured power in each output waveguide obviously scales with incident power so the tilt angle sensitivity will ultimately be limited by the incident power and signal-to-noise ratio of the photodetectors. The minimum power coupled into either grating in these devices is never less than 5% of the peak coupled power within the ±2° tilt angle operating range. Since wavefront angle measurements are carried out at relatively low speeds of at most a few kHz, highly sensitive photodetectors can be used.

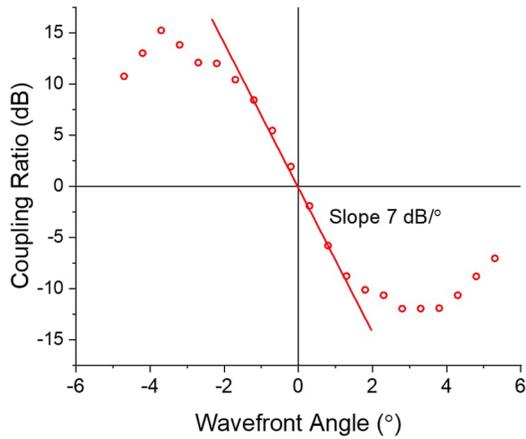

Fig. 2. The measured coupling ratio R1/R2 variation with incident wavefront angle for a 2 nm wide incident wavelength band from λ = 1525 to 1527 nm, for the ratio-based wavefront sensor configuration shown in Fig. 1(a).

The interferometric wavefront sensor configuration of Fig. 1 (b) was measured with the incident plane wave illuminating all four gratings. In these measurements the incident wavelength band was from λ = 1520 nm to 1525 nm, centered around the coupling efficiency maximum at normal incidence in these gratings. The four output signals shown in Fig. 3 were measured as the incident angle was varied. Outputs F1 and F3 provide an angle-dependent interference pattern between light intercepted by two gratings in the top row and bottom row of the sensor array, respectively. The resulting fringe patterns can be used to infer the wavefront tilt along the x-direction. If the incident wavefront is a perfect plane wave and all optical waveguide paths are identical, the F1 and F3 fringes would be identical for all angles. Similarly, outputs F2 and F4 are the interference patterns generated between the two left column gratings and the two right column gratings and provide a measure of the wavefront tilt along the y-direction. The interference signals will have the form

$$I(\theta) = I_0 \cdot A(\theta)[1 + \cos(D \cdot F(\theta) + \delta)] \quad (1)$$

where $I_0$ is the incident light intensity. As before, $A(\theta)$ is the angular dependence of grating coupling efficiency for an incoming plane wave. The effective grating separation D is the distance between the two apparent phase origins for light coupled into the waveguides by the two gratings. The value of D is different from geometrical grating center separation, and will vary slightly with incident angle and wavelength due to the corresponding variation of local grating diffraction strength. The multiplier $F(\theta) = (2\pi/\lambda) \cdot \sin(\theta)$ is the geometrical proportionality between D and the phase delay between the light incident on the two gratings. The term δ is an adjustable phase constant that is determined by the interferometer configuration and also any differences in optical path length connecting the two gratings and the Y-junctions. For the nominal design, δ = 0 for outputs F2 and F4, while δ = π for outputs F1 and F3 since the incident field arrives at the F1 and F3 combiners in antiphase at normal incidence.

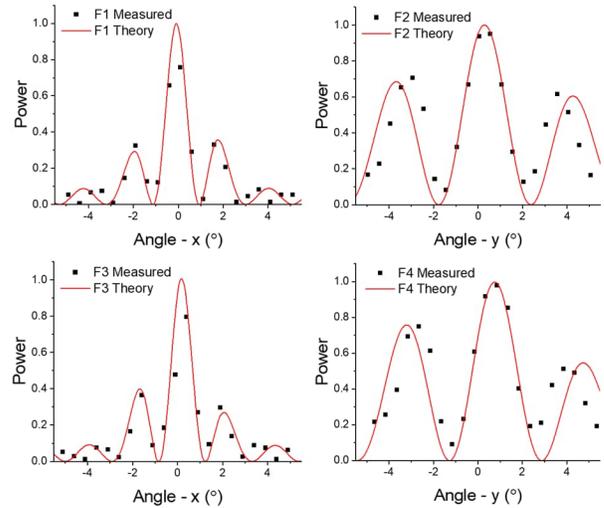

Fig. 3. The measured and calculated interferometric output signals for the wavefront sensor configuration of Fig. 1(b). For the output signals F1 and F3 the incident angle is varied along the x-axis in Fig. 1 while for outputs F2 and F3 the angle is varied along the y-direction.

Equation 1 was used to generate the theory curves in Fig. 3. For each curve the coefficients D and δ were adjusted to fit the central maximum lobe of the fringe pattern, but for simplicity both D and δ were approximated as being independent of incident angle. For the F1 and F3 model curves the effective grating separation used was D = 42 μm, while for F2 and F4 the separation was D = 21 μm. A Gaussian coupling envelope $A(\theta)$ with the as measured FWHM of 4° along the x-axis and FWHM of 14° along the y-axis was found to give a reasonable fit for all measurements. The qualitative comparison between theory and experiment is good, but at larger angles the measured fringe period becomes smaller than theory, due to the small increase in the effective grating separation D with angle. Each measured interference pattern is also offset from θ = 0°, because of unintended optical pathlength differences between interferometer arms, due to variations in fabricated waveguide and grating dimensions across the chip. These variations are accommodated in the model curves by adjusting δ to fit the measured fringe pattern. The waveguides from grating to combiner are up to 1 mm long, as shown in Fig. 1(b). Even an optical pathlength difference of only 0.1% will shift the observed fringe pattern by more than a full 2π cycle, and this must be taken into account in sensor calibration. The variation of wavefront tilt or phase over the sensor area can be inferred from the measured interference signals of Fig. 3. The fringe periods correspond to a 2° tilt angle variation along the x-direction for F1 and F3 outputs, and a 4° tilt along y for F2 and F4.

Two wavefront sensor designs based on Si photonic surface grating couplers have been described and characterized. The first design uses the ratio of coupled powers from two oppositely oriented grating couplers as the measure of local wavefront tilt, and has the advantage of simplicity. For the current design a measured tilt response of 7dB/° is obtained. The sensitivity can be increased by decreasing the angular width of the far-field coupling envelope, for example by using large area gratings. However, the tilt angle resolution will ultimately depend on the incident light power and the detector noise floor.

The second sensor configuration uses an arrangement of four identical gratings connected by an interferometric waveguide circuit. The four output fringe patterns can be used to monitor time dependent changes in local wavefront phase across the grating array, and therefore variations in tilt along the x and y directions. However, in the static case measuring any single output leaves the absolute value of phase and hence tilt angle unknown because the intensity of the incident beam is not known. Fortunately, this phase ambiguity can be removed (to mod($\pi$)) if the ratio of two parallel outputs such as F1 an F3 are measured simultaneously and compared, provided that the phase offset $\delta$ in Eq. 1 is different for the two output waveguides. With pre-calibration of each interferometer output, or by using active tuning on the waveguide arms, complete phase information (mod($\pi$)) can then be extracted from the four outputs. Given typical fabrication variations, the value of $\delta$ is unlikely to be precisely zero for any given device, but its value is unpredictable. This uncertainty can be removed by adding an active phase shifter to the output waveguides. The relative phase offset $\delta$ of beams arriving at each combiner can then be adjusted to any chosen set point. Furthermore, by actively scanning the phase delay of one waveguide of each output pair, the interference pattern can be sampled over a $2\pi$ phase range in real time, thereby removing any ambiguity in phase of light arriving at each grating. The sensitivity to wavefront tilt in the interferometric device of Fig. 1(c) can be increased by using a larger grating separation.

Either of the sensors shown in Fig. 1(b) and Fig. 1(c) can form a single element of a full wavefront monitoring system. Wavefront distortions over a wide area can be monitored by combining several of either these sensor elements into an array across a chip. The wavefront sampling density and number of sensor elements in such an array will be constrained by the need to route the output waveguides from each grating pair to the edge of the chip and hence to the photodetectors. Routing is straightforward for wavefront sensor applications requiring a relatively small number of array elements, such as in free space communication links, but may be more challenging when many hundreds of sensor elements in a close tiling over a 2-dimensional area are required as may be the case for imaging with large telescopes. Both grating sensor element designs have a much smaller area than the typical unit cell of a Schack-Hartmann sensor [1,2], which must be well over 100 $\mu$m across to accommodate one micro-lens and at least four pixels of an imaging array. The sensor size can be further reduced by using smaller gratings, albeit with the penalty that the power captured by the gratings will scale with grating area. Since these Si photonic sensor do not require an imaging array, photodetectors optimized for power sensitivity and for wavelengths where imaging arrays are not readily available can be used. Each grating sensor can only monitor one polarization component of the incident light, since the gratings couple TE polarized light into the waveguides. Although atmospheric wavefront distortion is usually not polarization dependent, if needed both orthogonal polarization components may be monitored by using two interleaved sets of sensor elements oriented at 90° to each other, or by using gratings with polarization diversity coupling [8].

The enabling component of these integrated optic wavefront sensors is the surface grating coupler optimized for near normal incidence. In this initial demonstration a vertical grating coupler designed for conventional chip to single mode fiber coupling has been used. In future work wavefront sensor designs could be enhanced using grating couplers optimized for free-space to waveguide coupling, with either smaller or larger coupling areas [9], and polarization diversity [8].